# Microresonator-Based Comb Generation without an External Laser Source

Adrea R. Johnson[1], Yoshitomo Okawachi[1], Michael R. E. Lamont[1,2,3], Jacob S. Levy[2], Michal Lipson[2,3], and Alexander L. Gaeta[1,3]

[1]*School of Applied & Engineering Physics, Cornell University, Ithaca, NY 14853*
[2]*School of Electrical and Computer Engineering, Cornell University, Ithaca, NY 14853*
[3]*Kavli Institute at Cornell for Nanoscale Science, Cornell University, Ithaca, NY 14853*

**Recent developments demonstrate that parametric four-wave mixing (FWM) in high-$Q$ microresonators is a highly promising and effective approach for optical frequency comb generation[1-10], with applications including spectroscopy, optical clocks, arbitrary waveform generation, frequency metrology, and astronomical spectrograph calibration[1,11-13]. Each of these microresonator platforms utilizes a scheme in which the system is pumped by a single-frequency laser at a cavity resonance. This scheme requires tuning of the pump wavelength into resonance and the generated comb is susceptible to fluctuations in pump power or frequency which can disrupt the soft thermal lock[14] and comb generation. We demonstrate a novel fiber-microresonator dual-cavity architecture that preferentially oscillates at modes of the microresonator due to its high density of states and generates robust and broadband combs (> 900 nm) without an external pump laser. Such a scheme could greatly simplify the comb generation process and allow for a fully-integrated chip-scale source with an on-chip amplifier[15].**

To date, frequency comb generation in microresonators requires tuning an external cw laser into a resonance of the microresonator. As pump power is coupled into the microresonator, thermal effects shift the resonance to higher wavelengths, creating a soft thermal lock between the cavity resonance and the pump laser[14]. When the intracavity power exceeds the threshold for parametric oscillation, cascaded four-wave mixing and higher-order four-wave mixing processes occur, resulting in comb generation. Fluctuations in the frequency or power of the pump laser can detune the pump out of the cavity resonance, disrupting comb generation. These limitations can be bypassed using the drop-port of the microresonator as shown by Peccianti *et al.*[16] with a hydex microring imbedded in a fiber cavity pumped with an erbium-doped fiber amplifier (EDFA). Through filter-driven four-wave mixing, they demonstrate a high-repetition-rate pulse source with a 60-nm bandwidth and a 200-GHz repetition rate. In this case, the drop-port acts as a filter, seeding the EDFA with wavelengths that correspond to resonances of the microresonator, similar to the recent work done by Cholan, *et al.*[17] for realization of a multiple wavelength source.

Here, we explore a dual-cavity architecture where a single bus waveguide (through-port), which forms part of the external fiber cavity, is coupled to the microresonator (Fig. 1). Due to the higher density of states within the microresonator as compared to the fiber cavity, preferential emission occurs at the microresonator modes resulting in lasing and parametric comb generation defined by the microresonator. This through-port configuration is effectively analogous to microresonator-based comb generation using a continuous wave (cw) pump laser. Additionally, since only a single bus waveguide is coupled to the microresonator, the system operates with reduced coupling losses and allows for higher power efficiency. Through the use of this dual-cavity design, we achieve broadband comb generation spanning more than 900 nm. Key advantages of this scheme are that it only requires a narrow-band optical amplifier, as opposed to a stabilized single-frequency laser, and it eliminates the need to pump at and tune to a resonance wavelength which eliminates pump frequency fluctuations that can shift the pump out of resonance and disrupt comb generation.

Our dual-cavity-based scheme for comb generation is shown in Fig. 1. Amplified spontaneous emission (ASE) from an EDFA is coupled into a silicon-nitride microresonator. A fiber polarization controller (FPC) allows for adjustment of the polarization of light coupled to the microresonator, which is a critical issue for cw lasing of the external fiber cavity and comb generation. A polarizer is placed at the output to select quasi-TE polarization. After the microresonator, a 3-dB coupler is used for output coupling. The remaining light in the external cavity is passed through a fixed bandpass filter (1553.83-1560.1 nm) and fed back into the EDFA. The bandpass filter allows for control of the spectral region that experiences round-trip EDFA gain. The waveguide cross section of 725 nm by 1600 nm of the silicon-nitride microresonator allows for a broad region of anomalous dispersion for broadband comb generation. We pump a 230-GHz free spectral range (FSR) microresonator with 2.17 W of EDFA power, and the resulting comb spectrum shown in Fig. 2 spans 900 nm (94 THz). We confirm that comb generation in this system is similar to that using an external cw pump laser by utilizing a 1.1-nm bandpass filter whose bandwidth is narrower than the FSR of the microresonator (**see Supplementary Information**). This restricts the amplifier bandwidth to a narrow region surrounding a single microresonator resonance and indicates that a single lasing peak is responsible for the generation of the comb. Additionally, we have achieved comb generation at an FSR of 80 GHz, with a bandwidth of 730 nm using 2 W of EDFA power (**see Supplementary Information**).

Our system achieves comb generation through a different mechanism as compared to the filter-driven FWM process[16] utilizing the drop-port of the microresonator. In the drop-port configuration, the microresonator effectively acts as a filter which selects the wavelengths that correspond to resonances of the microresonator for amplification in the EDFA. In our system, depending upon the state of the input polarization, lasing of the dual-cavity can be achieved at modes corresponding to the external fiber cavity or at modes of the microresonator. Initially, it may be counterintuitive that the cavity lases at microresonator resonances, where the round-trip cavity losses are higher. However, the mechanism can be understood in terms of the transition probability as dictated by Fermi's Golden Rule[18,19]. Below the lasing threshold, the probability of the system emitting at either a microresonator mode or a mode of the external fiber cavity is directly proportional to the density of available states at that frequency. The enhancement of the density of states in a cavity is given by the Purcell factor $F_p = (3\lambda^3/4\pi^2)(Q/V)$, where $Q$ is the quality factor and $V$ is the mode volume[20]. While the loaded $Q$-factor of the microresonator, when considering the cold-cavity modes, is lower than that of the external cavity by approximately a factor of 100, the mode volume of the microresonator cavity is lower by a factor of $10^6$, in which case the microresonator resonances provide a higher density of states and thus a higher transition probability to the frequencies corresponding to the microresonator resonance. Since the threshold power for lasing contains the same $V/Q$ factor[21], the modes corresponding to the microresonator cavity reach the oscillation threshold before the external fiber cavity. With sufficient power buildup, the microresonator cavity reaches the oscillation threshold, at which point, frequency dependent EDFA gain will preferentially amplify the oscillating mode(s) to the point where it can serve as a pump for comb generation. For comb generation, the polarization is adjusted to quasi-TE, as the coupling between the microresonator and the bus waveguide is optimized for the TE mode. When the polarization is adjusted away from quasi-TE, lasing occurs in modes corresponding to the external fiber cavity.

We simulate the lasing behavior in our system based on the coupled ring resonator model[22]. Figure 3 shows the time dynamics of the output of the dual-cavity system for various external fiber cavity round-trips ($T_R$) as the system reaches steady state. Figure 3a shows the system initially pumped with ASE. Initially, as the power builds up, the spectrum shows dips corresponding to resonances of the microresonator (Fig. 3b). Due to its high $Q$-factor and high modal confinement, the microresonator intracavity power continues to build with each successive round trip and reaches threshold for lasing before the external cavity. The lasing mode in the microresonator builds up further and dominates the external cavity modes resulting in the overall dual-cavity system lasing at modes corresponding to the microresonator (Fig. 3c-e). Alternatively, by increasing the cavity loss which decreases the $Q$-factor of the microresonator cavity, the dual-cavity system can be shown to lase at modes of the external fiber cavity away from the microresonator resonances.

We experimentally demonstrate lasing at modes of the microresonator and external fiber cavities using the dual-cavity setup minus the bandpass filter. We pump with low EDFA power such that the system is near the lasing threshold, and adjust the polarization to select the lasing wavelength of the system. For TE polarization, Fig. 4a illustrates lasing at modes of the microresonator for a range of power levels coupled into the bus waveguide. Dips in the optical spectrum correspond to TM resonances of the microresonator. Selecting TM polarization leads to lasing at modes of the external cavity away from the microresonator resonances (see Fig. 4b). The modulations in the spectrum result from a Fabry-Perot effect caused by reflections from the facets of the bus waveguide. Slight dips in the spectrum at 1542.5 nm and 1544.3 nm correspond to losses due to the TE resonances of the microresonator. The lasing wavelength of the dual-cavity system is therefore selected through the state of the input polarization (**see Supplementary Information**).

Finally, to understand the spectral properties of our dual-comb system, we perform simultaneous RF and optical spectral measurements of the microresonator output. By adjusting the polarization we allow the external fiber cavity to lase without interaction with the resonator (see Fig. 5a). The output from the bus waveguide is then sent to a 10-GHz photodiode and the RF signal is measured using an RF spectrum analyzer. The detected RF beatnote is 4.1 MHz (Fig. 5c), which corresponds to the mode beating of the external cavity, which occurs since the 1-GHz linewidth of the microresonator mode supports multiple modes of the external cavity. Thus, unlike the previous demonstration using a cw laser, the effective pump in this case is multi-mode. The intracavity polarization is rotated to quasi-TE for comb generation, and the entire comb output is sent to the photodiode for RF characterization. When a state of steady comb generation is reached (Fig. 5b), despite the multi-mode pump, a steady state of comb generation can be reached in which the RF amplitude noise drops by 40 dB (Fig. 5d). We expect that complete stabilization of the comb can be achieved through FSR control of both the microresonator and external fiber cavity.

In summary, we have demonstrated a dual-cavity design to produce broadband parametric frequency comb generation spanning 94 THz without the need of an external pump laser. As a result of the higher density of states within the microresonator as compared to the fiber cavity, the system automatically operates at a microresonator resonance which eliminates the need to carefully tune a cw laser into a resonance and results in comb generation that is inherently robust. Future designs could allow for the inclusion of an on-chip amplifier into the dual-cavity design to realize a fully-integrated chip-scale ultra-broadband frequency comb source.

**Figure 1 | General scheme.** Figure shows experimental setup for comb generation. The system is based on a dual-cavity design consisting of the silicon nitride microresonator and an external fiber cavity.

**Figure 2 | Dual-cavity comb spectrum.** Parametric frequency comb spectrum generated from a 230-GHz FSR silicon-nitride microresonator.

**Figure 3 | Simulation of lasing behavior in dual cavity.** Spectra illustrates the time dynamics of the dual-cavity intracavity power at various round-trips ($T_R$) in the fiber cavity as the system reaches steady state. **a,** Initial state is seeded with

noise. **b,** Power in the fiber modes exceeds the power microresonator modes. **c-d,** Buildup of intracavity power in the microresonator. **e,** Lasing of the system at modes of the microresonator.

**Figure 4 | Lasing modes of dual-cavity system.** Output spectrum of the dual-cavity system for a range of power levels in the bus waveguide. **a,** Dual-cavity system lasing at modes of microresonator. **b,** Dual-cavity system lasing in an external fiber cavity mode away from a microresonator resonance.

**Figure 5 | RF characterization of dual-cavity-based comb. a,** Optical spectrum for cw lasing of the external cavity. **b,** Optical frequency comb spectrum. **c,** RF spectrum for cw lasing gives a 4.1 MHz beatnote **d,** RF spectrum for optical frequency comb shows the near absence of an amplitude beatnote.

**Acknowledgment**

We acknowledge support from Defense Advanced Research Projects Agency (DARPA) via the QuASAR program and the Air-Force Office of Scientific Research under grant FA9550-12-1-0377. This work was performed in part at the Cornell Nano-Scale Facility, a member of the National Nanotechnology Infrastructure Network, which is supported by the National Science Foundation (NSF) (grant ECS-0335765). We also acknowledge useful discussions with I. H. Agha, K. Saha, R. Salem, and Y. H. Wen.

**Author contributions**

A.R.J. prepared the manuscript in discussion with all authors. A.R.J. and Y.O. designed and performed the experiments. Y.O. and M.R.E.L. performed numerical modeling. J.S.L fabricated the devices. M.L. and A.L.G. supervised the project.

**Additional information**

The authors declare no competing financial interest.

**Corresponding author**

Correspondence: Author to whom all correspondence should be directed.

Alexander L. Gaeta – a.gaeta@cornell.edu.

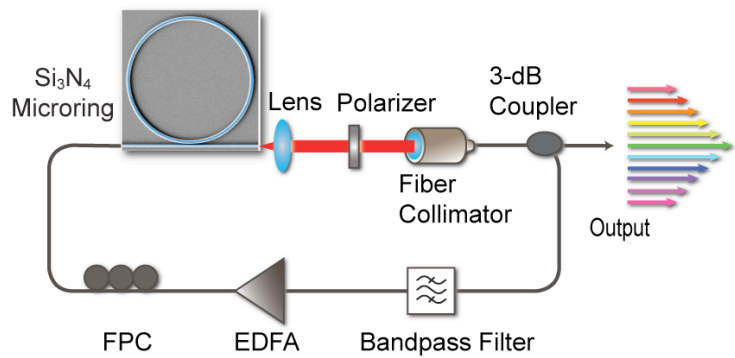

**Figure 1 | General scheme.** Figure shows experimental setup for comb generation. The system is based on a dual-cavity design consisting of the silicon nitride microresonator and an external fiber cavity.

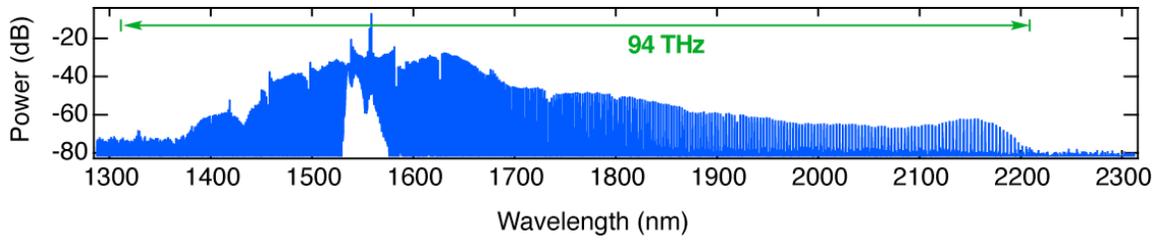

**Figure 2 | Dual-cavity comb spectrum.** Parametric frequency comb spectrum generated from a 230-GHz FSR silicon-nitride microresonator.

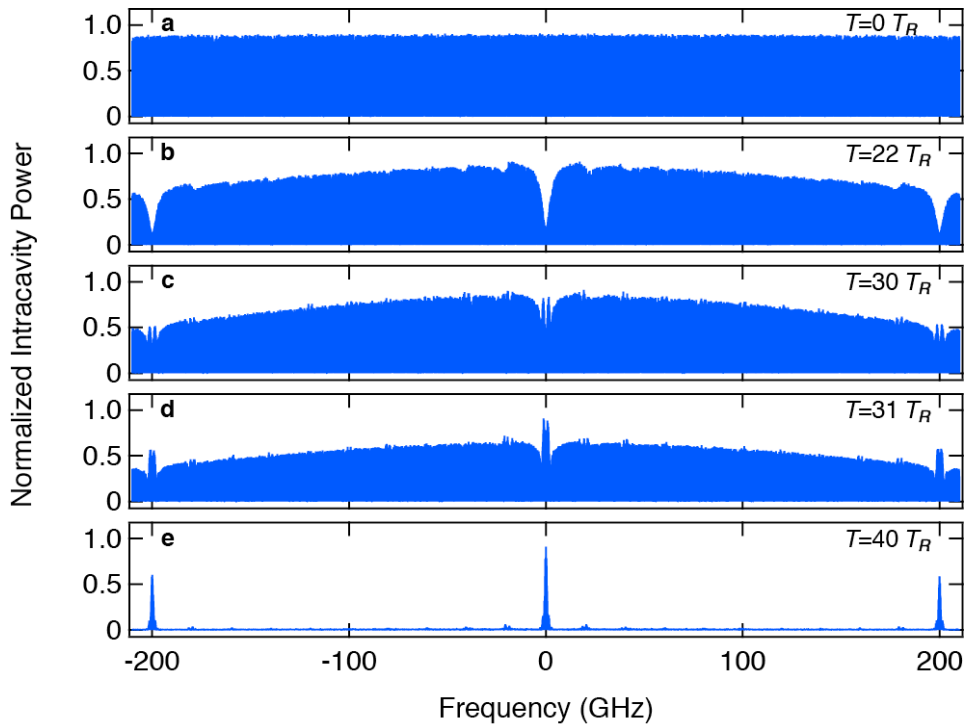

**Figure 3 | Simulation of lasing behavior in dual cavity.** Spectra illustrates the time dynamics of the dual-cavity intracavity power at various round-trips ($T_R$) in the fiber cavity as the system reaches steady state. **a,** Initial state is seeded with noise. **b,** Power in the fiber modes exceeds the power microresonator modes. **c-**

**d,** Buildup of intracavity power in the microresonator. **e,** Lasing of the system at modes of the microresonator.

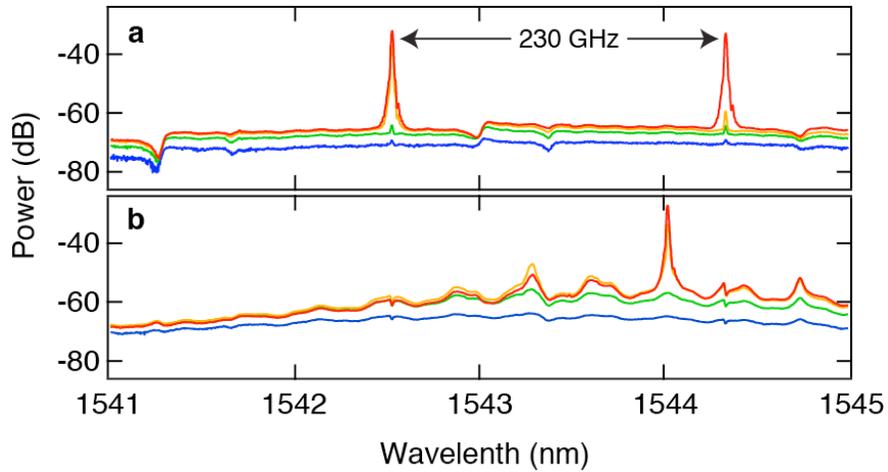

**Figure 4 | Oscillating modes of dual-cavity system.** Output spectrum of the dual-cavity system for a range of power levels in the bus waveguide. **a,** Dual-cavity system oscillating at modes of microresonator. **b,** Dual-cavity system oscillating in an external fiber cavity mode away from a microresonator resonance.

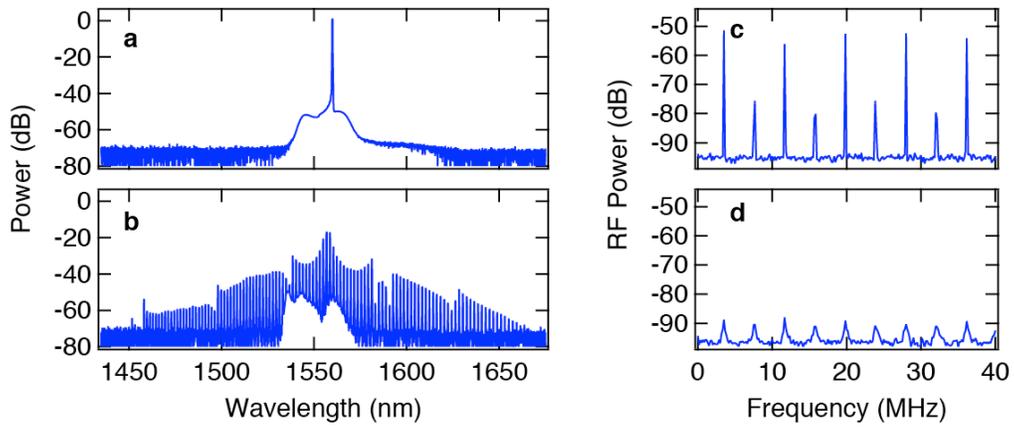

**Figure 5 | RF characterization of dual-cavity-based comb. a,** Optical spectrum for cw lasing of the external cavity without interaction with the microresonator. **b,** Optical frequency comb spectrum. **c,** RF spectrum for cw lasing gives a 4.1 MHz beatnote **d,** RF spectrum for optical frequency comb shows the near absence of an amplitude beatnote.